\documentclass[11pt,a4paper]{article}
\pdfoutput=1

\usepackage{jinstpub}
\usepackage{lineno}
\usepackage{indentfirst}
\usepackage{adjustbox}
\usepackage{multirow}
\usepackage{subfigure}
\usepackage{makecell}
\usepackage{siunitx}
\usepackage{enumerate}
\usepackage{graphicx}
\usepackage{amsmath}

\title{Development of a Monte Carlo tool for simulating electron transport in noble liquids}

\author[a,b]{Yijun~Xie}
\author[a,b,1]{and Yi~Wang\note{Corresponding author.}}

\affiliation[a]{Institute of High Energy Physics, Chinese Academy of Sciences\\Beijing 100049, China}
\affiliation[b]{University of Chinese Academy of Sciences\\Beijing 100049, China}

\emailAdd{wangyi90@ihep.ac.cn}

\abstract{
This study presents a Monte Carlo simulation tool for modeling the transportation processes of thermal electrons in noble liquids, specifically focusing on liquid argon and liquid xenon. 
The study aims to elucidate the microscopical mechanisms governing the drift and diffusion of electrons within the context of time projection chambers (TPCs), with detailed considerations of coherent electron-atom scattering and electric field force.
The simulation tool is implemented in the Geant4 framework, allowing for the exploration of electron transport parameters, including drift velocity, longitudinal diffusion coefficient, and transverse diffusion coefficient. 
The simulation is validated by comparing its results for drift velocity and diffusion coefficients with experimental measurements, revealing reasonable agreement in the low to moderate electric field ranges.
Discrepancies between the simulation and experimental measurements are discussed, emphasizing the necessity for enhanced cross-section calculations and high-precision sampling. 
Despite certain limitations, the simulation tool provides valuable insights into electron transport in noble liquids, establishing a foundation for future enhancements and applications in various research areas.
}

\keywords{Charge transport, multiplication and electroluminescence in rare gases and liquids, Detector modelling and simulations II (electric fields, charge transport, multiplication and induction, pulse formation, electron emission, etc), Noble liquid detectors (scintillation, ionization, double-phase), Simulation methods and programs}

\begin{document}

\maketitle
\flushbottom

\section{Motivation and introduction}
\label{sec:intro}

The technology of the noble liquid time projection chamber (TPC) plays a crucial role in high-energy physics experiments. 
It is extensively utilized in various experiments, including the search for weakly interacting massive particle (WIMP) dark matter~\cite{darkside50h, darkside50l, xenonnt, LZ, pandax4t}, the measurement of neutrino properties~\cite{protodune}, and the observation of neutrinoless double beta decay ($0\nu\beta\beta$)~\cite{exo200r}.
Understanding the transport of ionized electrons in noble liquids under electric fields is pivotal for TPC detectors, whether in single-phase (liquid only) or dual-phase (liquid and gas) configurations. 
This knowledge is essential for achieving high-resolution three-dimensional position reconstructions within the detector, for rejecting background noise from signals, and for comprehending detector properties.

Three key parameters characterize the electron transport in noble liquids under an electric field: the drift velocity $v_{e}$, the longitudinal diffusion coefficient $D_{L}$, and the transverse diffusion coefficient $D_{T}$. 
Measurements of these parameters in the region of interest have been conducted in both liquid xenon and liquid argon with various detector configurations, such as~\cite{exo200e, ICARUS}. 
However, a microscopic study elucidating the electron transportation processes to predict electron motion in noble liquids under different configurations is yet to be thoroughly explored. 
While tools like NEST (Noble Element Simulation Technique)~\cite{NEST} have features for simulating the motion of ionized electrons in xenon, and TRANSLATE (TRANSport in Liquid Argon of near-Thermal Electrons)~\cite{translate} is developed to simulate electron transport in liquid argon, there is worth in developing a tool that starts from first principles and can be applied to both liquid xenon and liquid argon.

This study aims to develop a Monte Carlo simulation tool for electron transportation processes in both liquid argon and xenon from first principles, employing random motion and collision of particles as the starting point.
The motion of electrons in noble liquids under an electric field, a process involving multiple elastic scattering, is described using mathematical tools such as stochastic differential equations, Langevin equations, and transition probabilities. 
Based on this, the microscopic mechanism of the electron transportation processes in noble liquids, as constructed by Van Hove et al.~\cite{Van_Hove}, is interpreted. 
Subsequently, a Monte Carlo sampling workflow is constructed and implemented under the Geant4 framework~\cite{geant4}. 
The simulation results, along with a comparison to experimental measurements, are presented in this work.

\section{Electron transport in noble liquids}

In this section, we will describe the electron transport process in noble liquids, based on the random motion and collisions among particles. 
The primary physical mechanism considered in this simulation tool is multiple elastic scattering between electrons and noble liquid atoms. 
We expounded on this process from a statistical perspective in section~\ref{sec:2.1}. 
Concurrently, each electron-atom scattering is treated as a coherent elastic scattering. 
We incorporate this microscopic process into the development of the simulation tool, as detailed in section~\ref{sec:2.2}.

\subsection{Multiple elastic scattering between electrons and atoms}
\label{sec:2.1}

In the context of the random motion of an ionized thermal electron in noble liquids, we employ a random force $\xi(t)$ to characterize the impact of frequent collisions among electrons and atoms, following stochastic thermal dynamics~\cite{ST}. 
The formal solution for the differential velocity of an electron is expressed as:
\begin{equation}
	\label{eq:Langevin_2}
    v(t) - v(0) = \int_{0}^{t} \xi(u)du.
\end{equation}

Here, a stochastic process $X(t)$ is introduced to represent the integral of the random force, facilitating subsequent statistical property calculations. 
$X(t)$ is subject to four hypotheses:

\begin{enumerate}[(1)]
    \item The increments of $X(t)$ are independent of each other:

Electrons in the noble liquid suffer from frequent random collisions, estimated to be on the order of $10^{12}$ collisions per second (as indicated in the total cross-section data presented in figure~\ref{fig:cross_section}, detailed description in section~\ref{sec:2.2}). 
We aim to mitigate the correlation of velocity changes between consecutive intervals, specifically, to flatten out the correlation between the interval $t_{n} - t_{n-1}$ and the previous interval $t_{n-1} - t_{n-2}$. 
This allows us to make the assumption that the increments of $X(t)$ are independent of each other. 
For any sequence of time points $0 \leqslant t_{1} \leqslant t_{2} \leqslant \ldots \leqslant t_{n}$, the differences $X(t_{n})-X(t_{n-1}), X(t_{n-1})-X(t_{n-2}), \ldots, X(t_{1})-X(0)$ are considered to be independent.

    \item Time translation invariance of the random force:

The time translation invariance of the random force allows us to choose the time origin arbitrarily. 
This implies that, for any fixed time interval $s > 0$, the random variables $X(t+s)-X(t)$ and $X(s)-X(0)$ follow identical distributions.

    \item Accelerated velocity should be finite:

The requirement for the accelerated velocity to be finite implies that any velocity-time sample curve must be continuous. 
Consequently, any sample curve of the stochastic process $X(t)$ is also continuous.

    \item The mean value of an increment $X(t)-X(0)$ equals zero:

The mean value of the increment $X(t)-X(0)$ is zero. 
Otherwise, the electron would gain energy from the liquid, violating the second law of thermodynamics.
\end{enumerate}

According to the Lévy-Itô decomposition theorem~\cite{Levy}, $X(t)$ must be a Brownian motion, which can be represented as $X(t)-X(0)\backsim\mathcal{N}(0,\zeta ^{2})$, where $\zeta$ represents the strength of the random force. 
Following Langevin's approach~\cite{Langevin}, a term $-\gamma v(t)$ is introduced to represent liquid resistance. 
Simultaneously, the strength of the electric field along the $z$-axis is denoted by $\mathcal{E}$. 
The Langevin equations describing the transportation processes of thermal electrons are as follows:
\begin{subequations}
  \label{eq:Langevin_3}
  \begin{align}
    \label{eq:Langevin_3:1}
    m\frac{dv_{T}(t)}{dt}     & = -\gamma_{T} v_{T}(t) + \zeta_{T} \frac{dB(t)}{dt}, \\
    \label{eq:Langevin_3:2}
    m\frac{dv_{L}(t)}{dt} & = \mathcal{E} e -\gamma_{L} v_{L}(t) + \zeta_{L} \frac{dB(t)}{dt}.
  \end{align}
\end{subequations}
Here, $B(t)$ denotes standard Brownian motion, $m$ denotes the mass of an electron, and $e$ represents the charge of an electron.
Equation~\eqref{eq:Langevin_3:1} describes transverse motion, while~\eqref{eq:Langevin_3:2} describes motion along the longitudinal direction. 

From equations~\eqref{eq:Langevin_3:1} and~\eqref{eq:Langevin_3:2}, the velocity of an electron can be derived as:
\begin{subequations}
  \label{eq:Langevin_4}
  \begin{align}
    \label{eq:Langevin_4:1}
    v_{T}(t)     & = v_{T}(0)e^{-\frac{\gamma_{T}}{m}t} + \frac{\zeta_{T}}{m} \int_{0}^{t} e^{-\frac{\gamma_{T}}{m}(t-s)}dB(s), \\
    \label{eq:Langevin_4:2}
    v_{L}(t) & = v_{L}(0)e^{-\frac{\gamma_{L}}{m}t} + \frac{\mathcal{E} e}{\gamma_{L}}(1-e^{-\frac{\gamma_{L}}{m}t}) + \frac{\zeta_{L}}{m} \int_{0}^{t} e^{-\frac{\gamma_{L}}{m}(t-s)}dB(s).
  \end{align}
\end{subequations}

From equation~\eqref{eq:Langevin_4:2}, the average velocity along the $z$-axis, once equilibrium is reached, is given by:
\begin{equation}
  \label{eq:Langevin_5}
  \left\langle v_{L} \right\rangle(\infty) = v_{d} = \frac{\mathcal{E} e}{\gamma_{L}},
\end{equation}
and the average velocity in the transverse direction, once equilibrium is reached, is:
\begin{equation}
  \label{eq:Langevin_6}
  \left\langle v_{T}^{2} \right\rangle(\infty) = \frac{\zeta_{T}^{2}}{2\gamma_{T}}.
\end{equation}
This result aligns with the physical expectation that ionized thermal electrons drift at a constant speed in noble liquids.

Regarding the dependence between position and time:
\begin{subequations}
  \label{eq:Langevin_7}
  \begin{align}
    \label{eq:Langevin_7:1}
    x_{T}(t)     & = x_{T}(0) + \frac{mv_{T}(0)}{\gamma_{T}}(1 - e^{-\frac{\gamma_{T}}{m}t}) + \frac{\zeta_{T}}{\gamma_{T}} \int_{0}^{t} [1-e^{-\frac{\gamma_{T}}{m}(t-s)}]dB(s),
    \\
    \label{eq:Langevin_7:2}
    x_{L}(t) & = x_{L}(0) + \frac{\mathcal{E} e}{\gamma_{L}}t + \frac{(mv_{L}(0)-\frac{\mathcal{E} e}{\gamma_{L}})}{\gamma_{L}}(1-e^{-\frac{\gamma_{L}}{m}}t)+\frac{\zeta_{L}}{\gamma_{L}} \int_{0}^{t} [1-e^{-\frac{\gamma_{L}}{m}(t-s)}]dB(s).
  \end{align}
\end{subequations}
The mean and variance of these functions can be expressed as:
\begin{subequations}
  \label{eq:Langevin_8}
  \begin{align}
    \label{eq:Langevin_8:1}
    \mu_{T}(t)        & = \left\langle x_{T}(t) \right\rangle = x_{T}(0) + \frac{mv_{T}(0)}{\gamma_{T}}(1 - e^{-\frac{\gamma_{T}}{m}t}),
    \\
    \label{eq:Langevin_8:2}
    \sigma_{T}^{2}(t) & = \left\langle (x_{T}(t)-\mu_{T}(t))^{2} \right\rangle = \frac{\zeta_{T}^{2}}{\gamma_{T}^{2}}(t-\frac{3m}{2\gamma_{T}}+\frac{2m}{\gamma_{T}}e^{-\frac{\gamma_{T}}{m}t}-\frac{m}{2\gamma_{T}}e^{-\frac{2\gamma_{T}}{m}t}),
  \end{align}
\end{subequations}
and
\begin{subequations}
  \label{eq:Langevin_9}
  \begin{align}
    \label{eq:Langevin_9:1}
    \mu_{L}(t)        & = x_{L}(0) + \frac{\mathcal{E} e}{\gamma_{L}}t + \frac{(mv_{L}(0)-\frac{\mathcal{E} e}{\gamma_{L}})}{\gamma_{L}}(1-e^{-\frac{\gamma_{L}}{m}}t),
    \\
    \label{eq:Langevin_9:2}
    \sigma_{L}^{2}(t) & = \left\langle (x_{L}(t)-\mu_{L}(t))^{2} \right\rangle = \frac{\zeta_{L}^{2}}{\gamma_{L}^{2}}(t-\frac{3m}{2\gamma_{L}}+\frac{2m}{\gamma_{L}}e^{-\frac{\gamma_{L}}{m}t}-\frac{m}{2\gamma_{L}}e^{-\frac{2\gamma_{L}}{m}t}).
  \end{align}
\end{subequations}

In accordance with the properties of Itô's integral~\cite{Ito} and transition probability, the probability of finding an electron at position $x$ at time $t$ given the initial position at time 0 is $x(0)$:
\begin{equation}
  \label{eq:Langevin_10}
  p(x,t | x(0),0) = \frac{1}{\sqrt{2\pi }\sigma_{\text{loc}}(t)} e^{-\frac{(x-\mu_{\text{loc}}(t))^2}{2\sigma_{\text{loc}}^{2}(t)}},
\end{equation}
where $\sigma_{\text{loc}}(t) = \sigma_{T}(t)~\text{or}~\sigma_{L}(t)$ and $\mu_{\text{loc}}(t) = \mu_{T}(t)~\text{or}~\mu_{L}(t)$. The comparison between this result and the solution of the diffusion equation gives:
\begin{equation}
  \label{eq:Langevin_11}
  D = \frac{\zeta_{\text{loc}}^{2}}{2\gamma_{\text{loc}}^{2}}.
\end{equation}

The form of equation~\eqref{eq:Langevin_10} is the core deduction of the Langevin equation, encapsulating all the essential physical information required for this study. 
It can be utilized to extract transportation parameters in Monte Carlo simulations from the position distribution of electrons in any independently conducted experiment.

\subsection{Microscopic mechanism: coherent electron-atom scattering}
\label{sec:2.2}

To model the transport processes of thermal electrons in noble liquids, we adopt a scattering perspective, to understand the microscopic mechanism involved. 
For thermal electrons traveling in noble liquids such as liquid argon or liquid xenon, the condition $\lambda_{e} > d_{\text{atom}}$ holds, where $\lambda_{e}$ is the de Broglie wavelength of an electron, and $d_{\text{atom}}$ is the average distance between atoms. 
This condition underscores the importance of considering the coherent effect in electron-atom scattering.

The incident and final wave vectors of an electron, denoted as $\boldsymbol{k_{0}}$ and $\boldsymbol{k}$ respectively, dictate the momentum transfer ($\boldsymbol{k}$) and energy transfer ($\omega$), given by:
\begin{subequations}
  \label{eq:momentum_and_energy_transfers}
  \begin{align}
    \label{eq:momentum_and_energy_transfers:1}
    \boldsymbol{\kappa} & = \boldsymbol{k_{0}} - \boldsymbol{k},
    \\
    \label{eq:momentum_and_energy_transfers:2}
    \omega              & = \frac{\hbar }{2m_{e}}(k_{0}^{2} - k^{2}).
  \end{align}
\end{subequations}

Van Hove's study in~\cite{Van_Hove} expresses the double-differential cross-section as:
\begin{equation}
  \label{eq:Van_Hove_XSec}
  \frac{d^{2}\sigma_{\text{coh}}}{d\Omega d\omega} = \sigma_{\text{eff}} S(\boldsymbol{\kappa},\omega).
\end{equation}
Here, $\sigma_{\text{eff}}$ is the effective cross-section derived from the effective potential in the liquid. 
Integrating over $\omega$, the differential cross-section per unit solid angle becomes:
\begin{equation}
  \label{eq:Van_Hove_XSec_2}
  \frac{d\sigma_{\text{coh}}}{d\Omega} = \int \frac{d^{2}\sigma_{\text{coh}}}{d\Omega d\omega} d\omega = \sigma_{\text{eff}} S(\boldsymbol{\kappa}),
\end{equation}
where the structure factor, $S(\boldsymbol{\kappa})$, is the Fourier transform of the radial distribution function $g(\boldsymbol{r})$~\cite{Lekner}:
\begin{subequations}
  \begin{align}
  \label{eq:S_k_g_r}
  S(\boldsymbol{\kappa}) & = 1 + n\int_{V}d\boldsymbol{r} e^{i\boldsymbol{\kappa}\cdot\boldsymbol{r}}(g(\boldsymbol{r})-1),
  \\
  \label{g_r}
  g(\boldsymbol{r}) & = \frac{1}{\rho}\left\langle \sum_{i\neq 1}\delta(\boldsymbol{r}-\boldsymbol{r_{i}})\right\rangle.
  \end{align}
\end{subequations}
Here, $g(\boldsymbol{r})$ represents the average number of liquid atoms found in the volume $d^{3}\boldsymbol{r}$ around the position $\boldsymbol{r}$.
$\rho$ stands for the density of the liquid.
Thus, $S(\boldsymbol{\kappa})$ is associated with the liquid's structure and can be measured through neutron scattering experiments.

\begin{figure}[htbp]
  \centering
  \subfigure[Argon]{\label{fig:sk_argon}
  \includegraphics[width=.49\textwidth,trim=1cm 0.5cm 0.5cm 0.5cm]{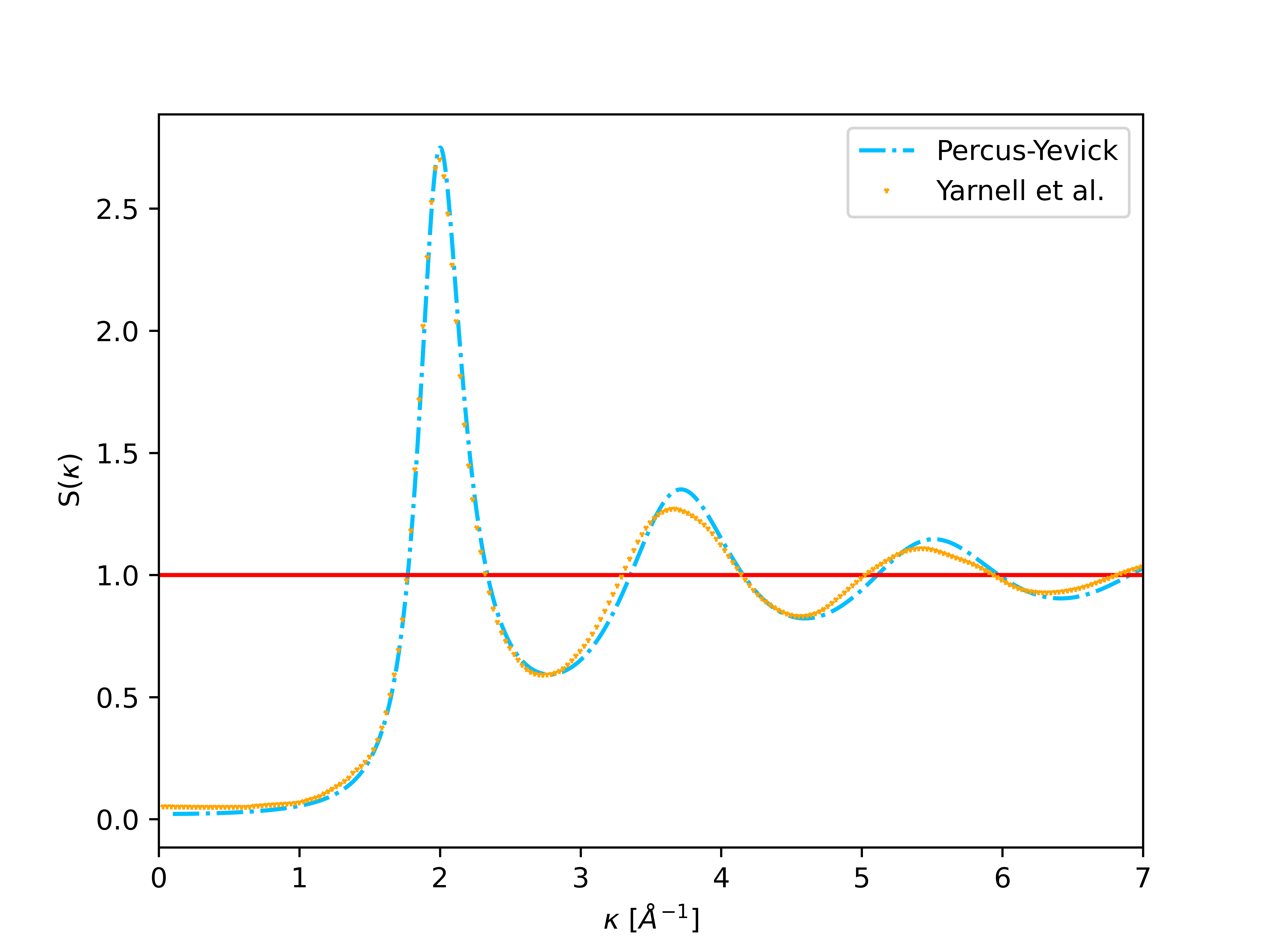}}
  \subfigure[Xenon]{\label{fig:sk_xenon}
  \includegraphics[width=.49\textwidth,trim=0.5cm 0.5cm 1cm 0.5cm]{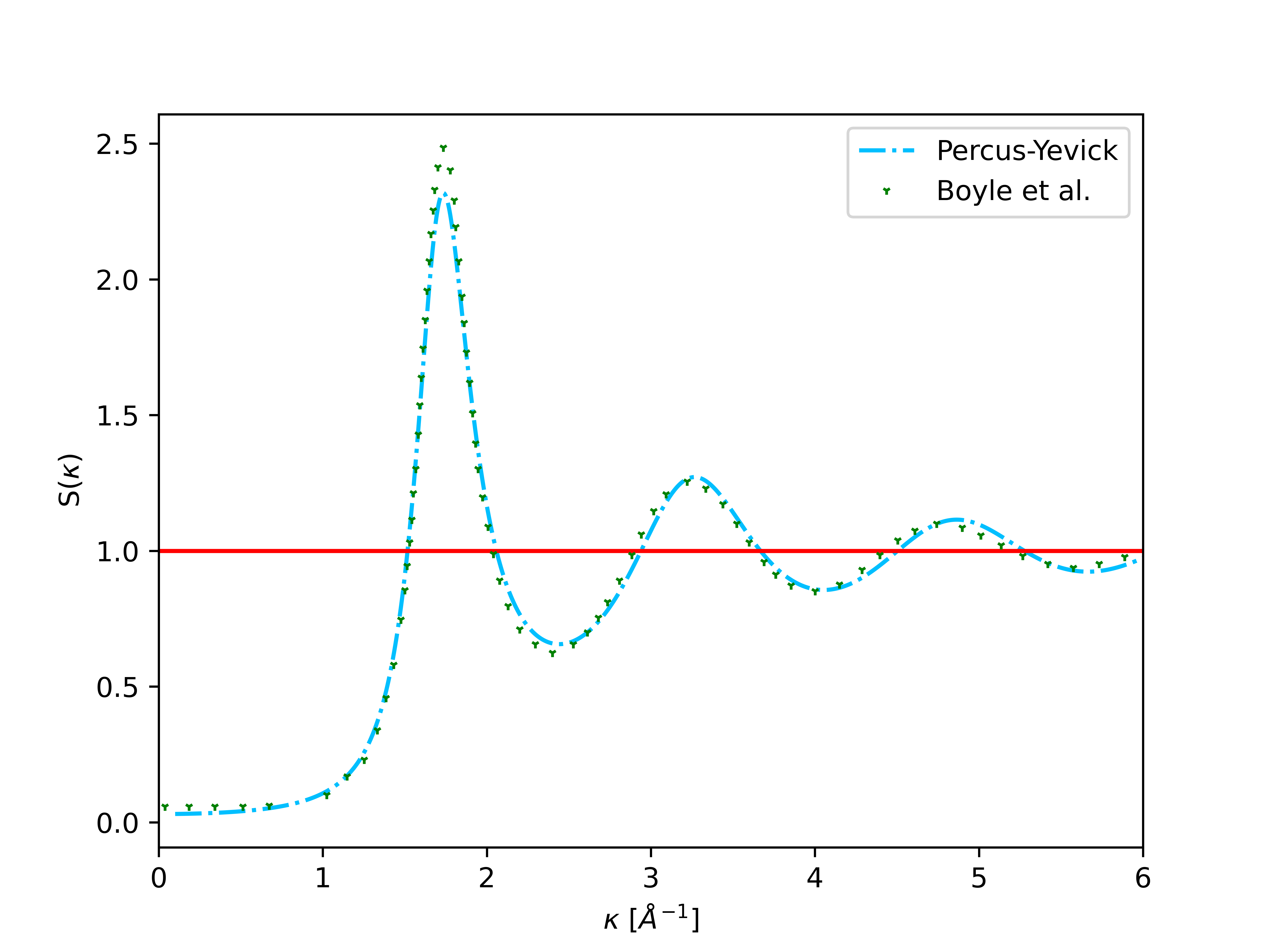}}
  \caption{\label{fig:S_k} The structure factor $S(\boldsymbol{\kappa})$ of noble liquids. The left plot displays the $S(\boldsymbol{\kappa})$ of liquid argon at 85~K. The blue dash-dotted line represents calculations based on a theoretical model from~\cite{Percus}, while the orange dotted line corresponds to measurements from a neutron scattering experiment conducted by Yarnell et al.~\cite{Yarnell}. The right plot displays the $S(\boldsymbol{\kappa})$ of liquid xenon at 165~K. The blue dash-dotted line represents calculations based on theoretical models from~\cite{Percus, Atrazhev_Liquid_Xenon}, while the green dotted line is derived from molecular dynamic simulations performed by Boyle et al.~\cite{Boyle2}. Both curves are expected to converge to 1, as indicated by the red line.}
\end{figure}

Figure~\ref{fig:S_k} illustrates $S(\boldsymbol{\kappa})$ in both liquid argon and liquid xenon. 
In Figure~\ref{fig:sk_argon}, the blue dash-dotted line represents the predicted $S(\boldsymbol{\kappa})$ from the Percus-Yevick integral equation, as described in~\cite{Percus}. 
The orange dotted line corresponds to measurements by Yarnell et al.~\cite{Yarnell} in a neutron scattering experiment. 
In Figure~\ref{fig:sk_xenon}, the blue dash-dotted line represents the predicted $S(\boldsymbol{\kappa})$ based on the theoretical model described in~\cite{Percus, Atrazhev_Liquid_Xenon}, while the green dotted line is derived from molecular dynamic simulations conducted by Boyle et al.~\cite{Boyle2}.
The $S(\boldsymbol{\kappa})$ serves as a crucial input for the Monte Carlo simulation tool, which will be introduced later.
It is noteworthy that, in noble liquids, although there is no long-range order as in a crystal, short-range order and correlations still exist. 
The oscillations in the structure factor reflect the periodicities and spatial arrangements of particles within a local region.

\begin{figure}[htbp]
  \centering
  \includegraphics[width=.8\textwidth]{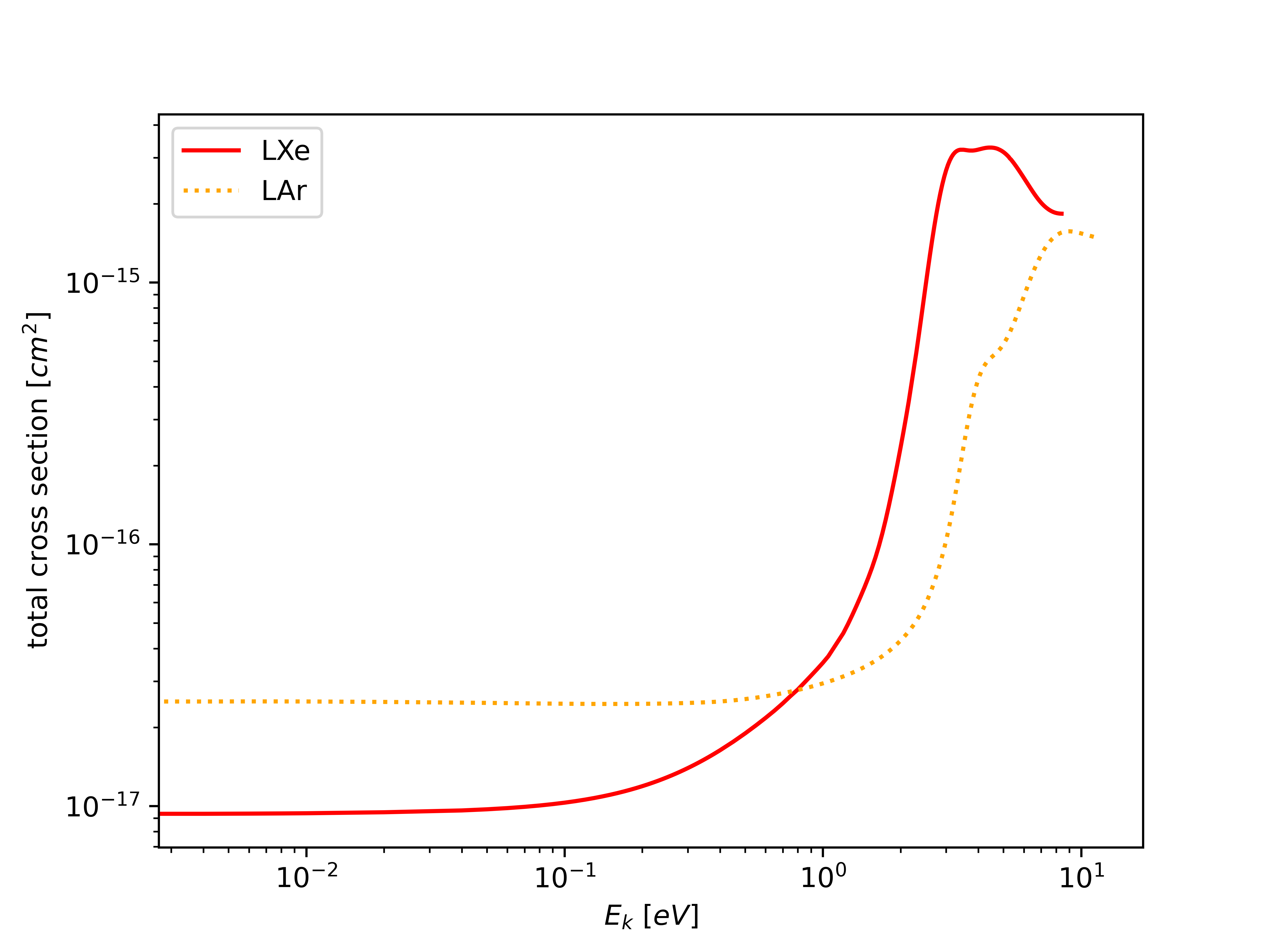}
  \caption{\label{fig:cross_section} Coherent electron-atom scattering cross-sections versus the kinetic energy of electrons for liquid argon (solid line) and liquid xenon (dashed line). The data is provided by Dr. Gregory James Boyle~\cite{Boyle,Boyle2}.}
\end{figure}

The coherent electron-atom scattering cross-sections of liquid argon and liquid xenon, as calculated by Boyle et al.~\cite{Boyle,Boyle2}, are shown in figure \ref{fig:cross_section}. 
This information serves as another input for the developed Monte Carlo simulation tool.

\section{Monte Carlo simulation setup}

The mean free path of an electron traveling in noble liquids, given by the product of the number density and total cross-section, $\Lambda = n_{\text{liq}}\sigma_{\text{coh}}$, must satisfy the condition:
\begin{equation}
  \label{eq:MFP}
  \Lambda > d_{\text{atom}}.
\end{equation}
Thus, the transportation processes of an electron can be separated into movement under an electric field and movement due to a scattering process. 
Concerning the motion of an electron under an electric field, it has been well defined in Geant4~\cite{geant4, G4_1, G4_2}. 
The "G4DormandPrince745" class is declared as the stepper to navigate electrons between the interaction points under an electric field.

For the scattering process of an electron traveling inside noble liquids, a Geant4 physics process, derived from the "G4VRestDiscreteProcess" class, was implemented to complete the Monte Carlo tool. 
Sampling of the scattering process is divided into angle change sampling and energy transfer sampling: $ P(\theta,\omega) = P(\omega\vert\theta)P(\theta) $. 
Concerning angle change sampling, equation~\eqref{eq:Van_Hove_XSec_2} is used to sample the change in direction $\Delta \theta$. Due to $ m_{e} \ll m_{\text{atom}} $, $\left\lvert \boldsymbol{\kappa} \right\rvert $ is estimated by $\left\lvert \boldsymbol{\kappa} \right\rvert = 2\left\lvert \boldsymbol{k_{0}} \right\rvert \sin(\frac{\Delta \theta}{2})$.

\begin{figure}[htbp]
  \centering
  \includegraphics[width=.49\textwidth,trim=1cm 0cm 0.5cm 1cm]{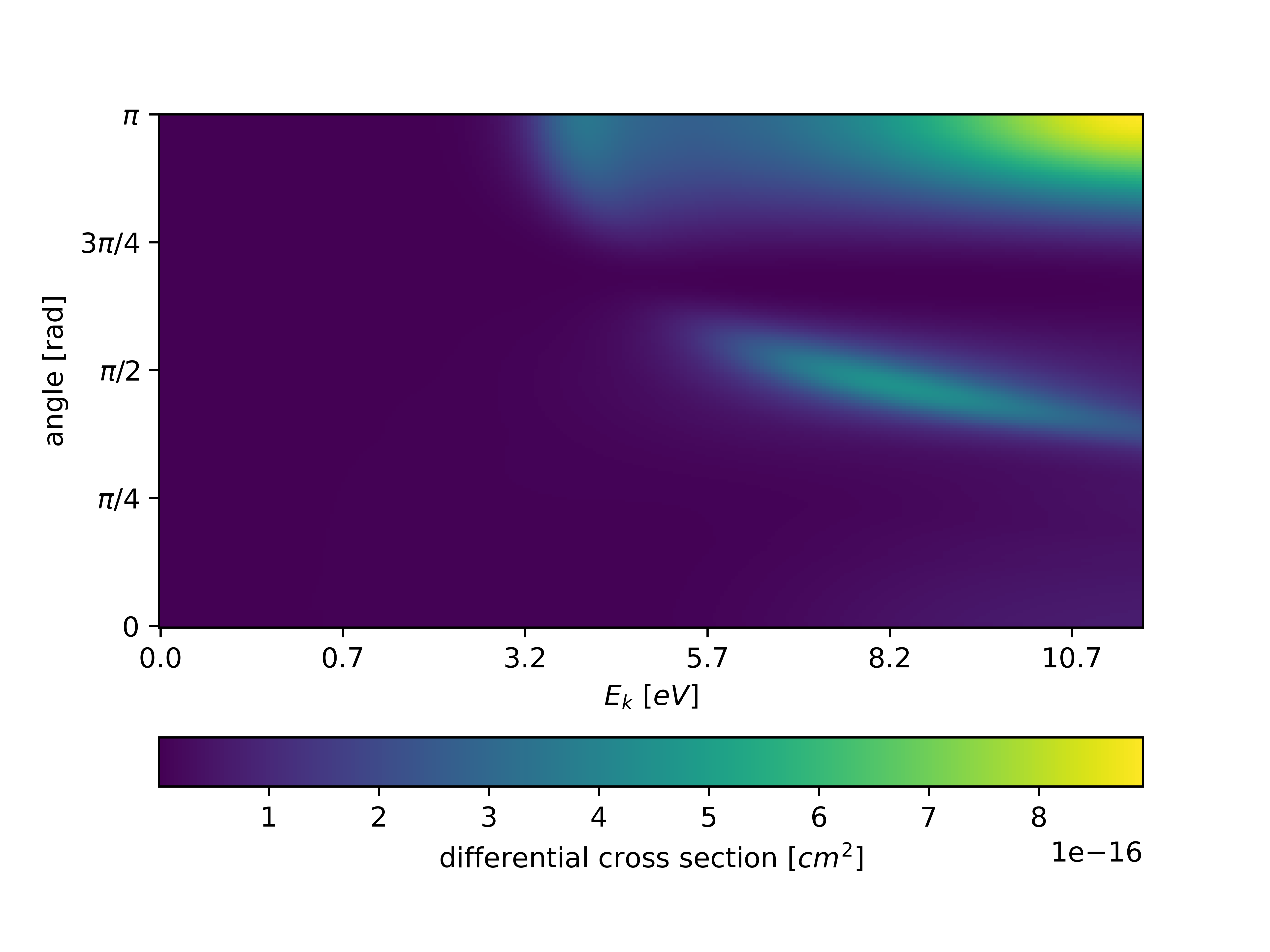}
  \includegraphics[width=.49\textwidth,trim=0.5cm 0cm 1cm 1cm]{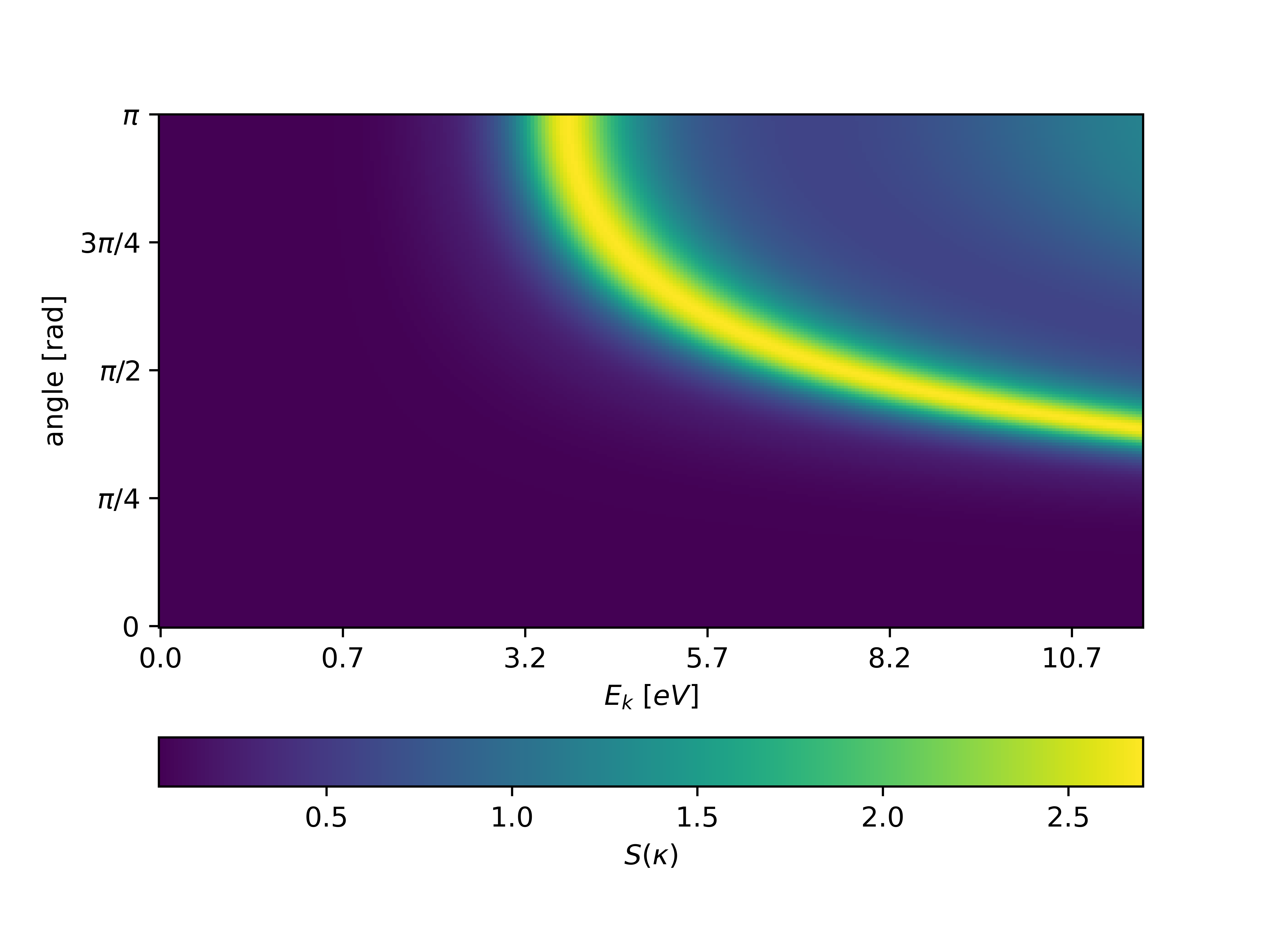}
  \qquad
  \includegraphics[width=.49\textwidth,trim=1cm 1cm 0.5cm 1cm]{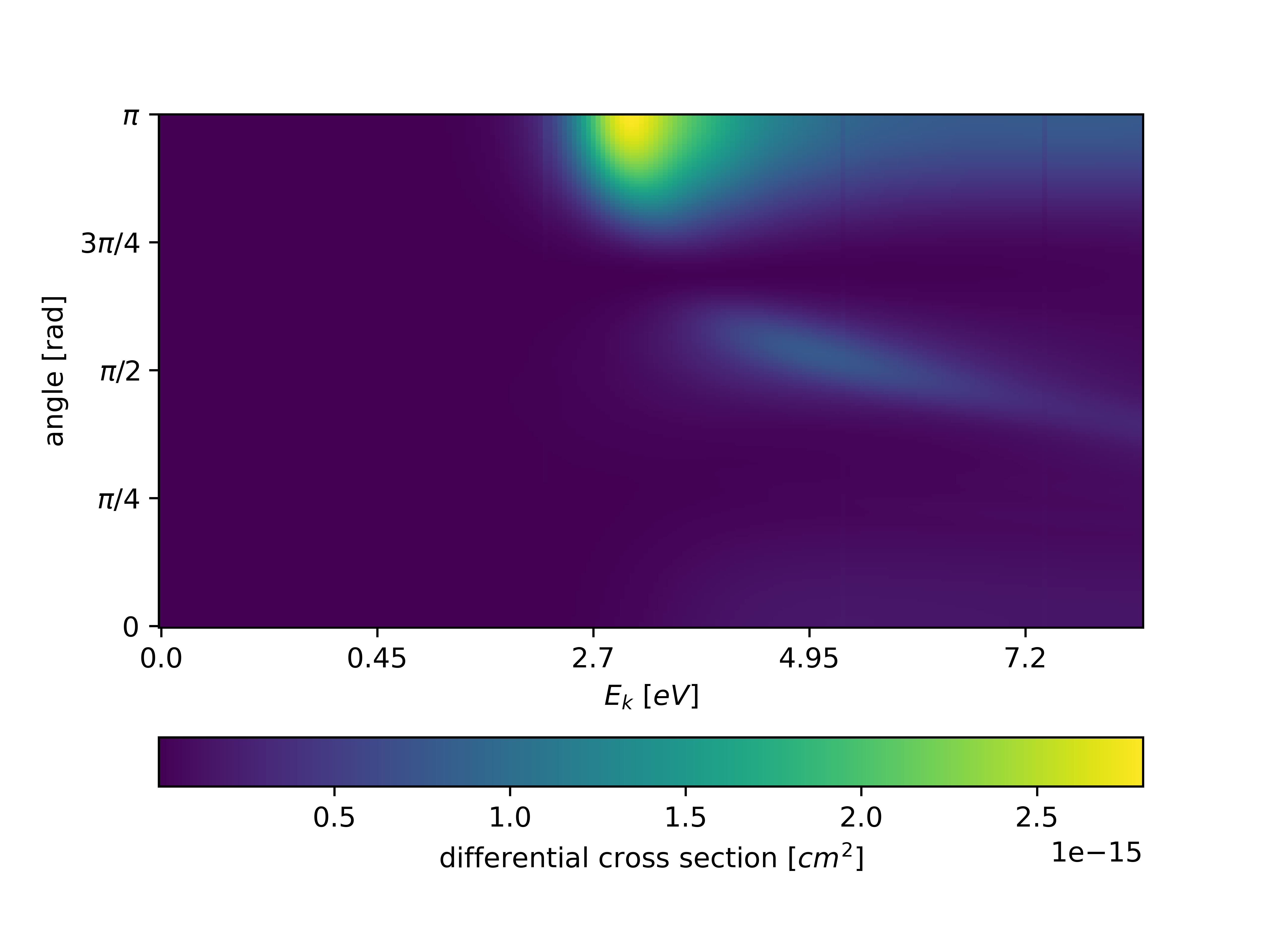}
  \includegraphics[width=.49\textwidth,trim=0.5cm 1cm 1cm 1cm]{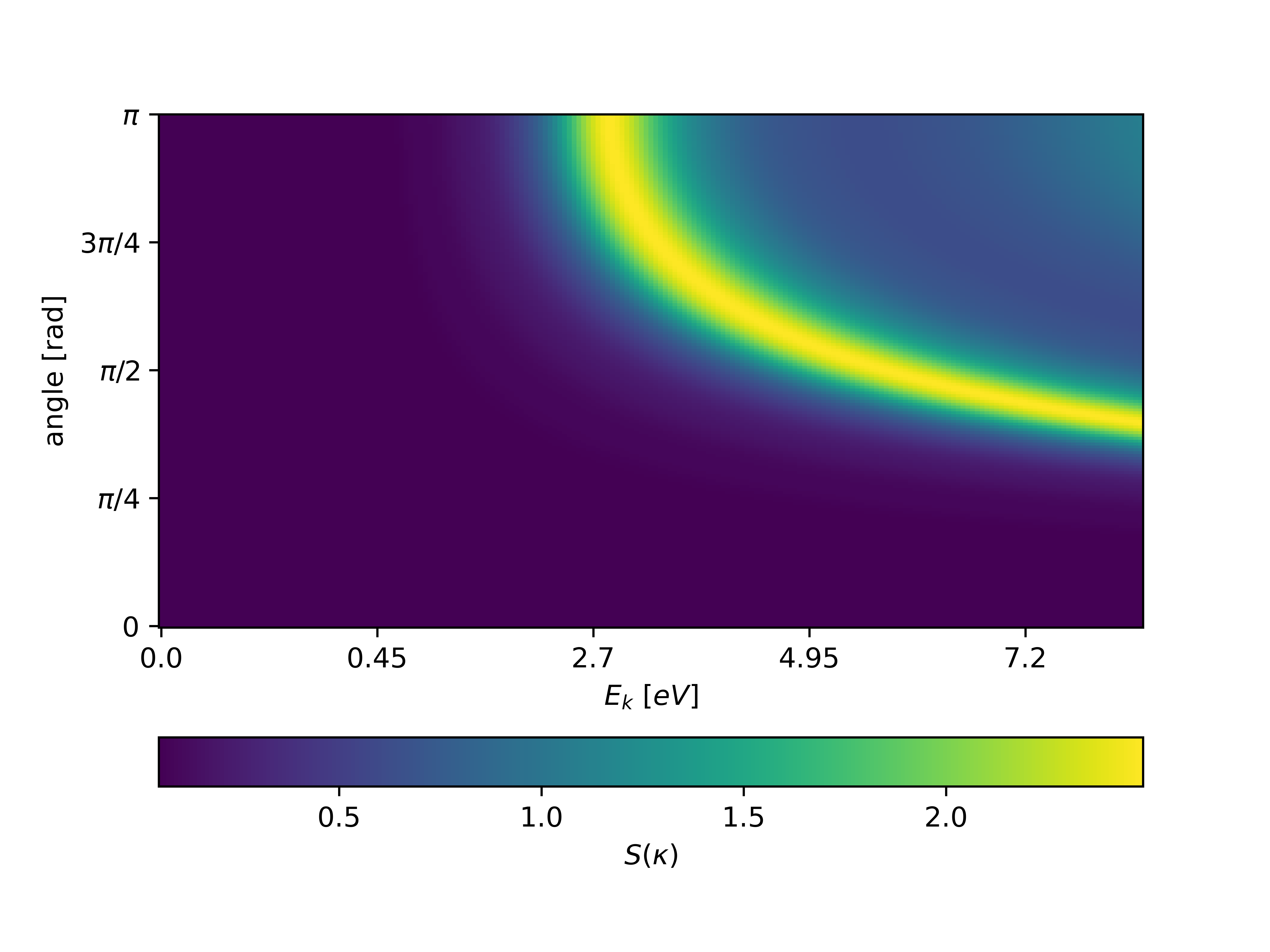}
  \caption{\label{fig:input_data} The differential cross-section, $\frac{d\sigma_{\text{coh}}}{d\Omega}$, and structure factor, $S(\boldsymbol{\kappa})$, used in the Monte Carlo simulation. The top two plots correspond to the results for liquid argon at 85~K, and the bottom two plots correspond to the results for liquid xenon at 165~K. The yellow band in $S(\boldsymbol{\kappa})$ corresponds to the first peak of in the $S(\boldsymbol{\kappa})$ plots shown in figure \ref{fig:S_k}. The enhancement of back-scattering direction implies that there is an upper limit to the drift velocity under strong electric fields.}
\end{figure}

Concerning energy transfer sampling, since the scattering takes place between the electron and the liquid system, by taking into account the ensemble average of the liquid system and referring to \cite{Rahman}, the statistical properties of $S(\boldsymbol{\kappa},\omega)$ in coherent scattering can be expressed as:
\begin{subequations}
  \label{eq:S_K_omega}
  \begin{align}
    \label{eq:S_K_omega:1}
    \int S(\boldsymbol{\kappa},\omega)d\omega            & = S(\boldsymbol{\kappa}), \\
    \label{eq:S_K_omega:2}
    \int \omega S(\boldsymbol{\kappa},\omega)d\omega     & = \frac{\hbar\left\lvert \boldsymbol{\kappa} \right\rvert^{2}}{2m_{\text{atom}}}, \\
    \label{eq:S_K_omega:3}
    \int \omega^{2} S(\boldsymbol{\kappa},\omega)d\omega & = 2k_{B}T\frac{\left\lvert \boldsymbol{\kappa} \right\rvert^{2}}{2m_{\text{atom}}} + O(\hbar^{2}).
  \end{align}
\end{subequations}
Here, $k_{B}$ stands for the Boltzmann constant, and $T$ stands for the temperature of the liquid. 
Combining~\eqref{eq:S_K_omega:1}, \eqref{eq:S_K_omega:2}, \eqref{eq:S_K_omega:3}, and $ P(\omega\vert\theta) = \frac{P(\theta,\omega)}{P(\theta)} = \frac{S(\boldsymbol{\kappa},\omega)}{S(\boldsymbol{\kappa})} $, we obtain the first-order moment of $P(\omega\vert\theta)$ in equation~\eqref{eq:P_omega_theta:1} and the second-order moment in equation~\eqref{eq:P_omega_theta:2}. 
\begin{subequations}
  \label{eq:P_omega_theta}
  \begin{align}
    \label{eq:P_omega_theta:1}
    M_{1} = \int \omega P(\omega\vert\theta)d\omega     & = \frac{\hbar\left\lvert \boldsymbol{\kappa} \right\rvert^{2}}{2m_{\text{atom}}S(\boldsymbol{\kappa})}, \\
    \label{eq:P_omega_theta:2}
    M_{2} = \int \omega^{2} P(\omega\vert\theta)d\omega & = 2k_{B}T\frac{\left\lvert \boldsymbol{\kappa} \right\rvert^{2}}{2m_{\text{atom}}S(\boldsymbol{\kappa})} + O(\hbar^{2}).
  \end{align}
\end{subequations}
With these two orders of moments, a normal distribution, $\mathcal{N}(\hbar M_{1},\hbar^{2} M_{2})$, is used to approximate the distribution of $P(\omega\vert\theta)$. 
Thus, with the differential cross-section defined in~\eqref{eq:Van_Hove_XSec_2}, shown in figure~\ref{fig:cross_section}, and $S(\boldsymbol{\kappa})$ designed in~\eqref{eq:S_k_g_r}, shown in figure~\ref{fig:S_k}, the input of this simulation tool is fully defined. 
Figure~\ref{fig:input_data} illustrates the input data of the simulation tool.

To set up the simulation, we define an active volume as a cylindrical region with a radius of 30 cm and a height of 60 cm. 
The electric field within this cylindrical region is specifically configured to be uniform along the $z$-axis, and its intensity is controlled using the "G4ElectroMagneticField" class. 
To initiate an event, a single electron is generated using the "G4ParticleGun" class.

Within each simulation process, electrons are initialized at the coordinates (0, 0, 0), and their x, y, z coordinates are recorded following drift for a specific duration. 
After exploring the process with various drift times, we have determined that 500~ns is the optimal duration for performing the simulations. 
This time interval is deemed sufficient for electrons to reach a stable motion state. 
The outcomes of the simulation are presented in the next section.

\section{Fitting results and discussions}

The position information of electrons drifting for 500~ns is utilized to directly extract velocity and longitudinal diffusion distributions, followed by fitting processes with Gaussian distributions.
Regarding the diffusion distribution in the transverse direction, considering the symmetry between the x and y coordinates, we extract the $D_{T}$ distribution along $r$, defined as $r = \sqrt{x^{2}(t) + y^{2}(t)}$, and fit it with the following equation:
\begin{equation}
  \label{eq:D_T_fit}
  f(r) = p_{0}\frac{r}{\sigma_{T}(t)}e^{-\frac{r^{2}}{2\sigma^{2}_{T}(t)}}.
\end{equation}

\begin{figure}[htbp]
  \centering
  \includegraphics[width=.51\textwidth,trim=0.2cm 0cm 0cm 0cm]{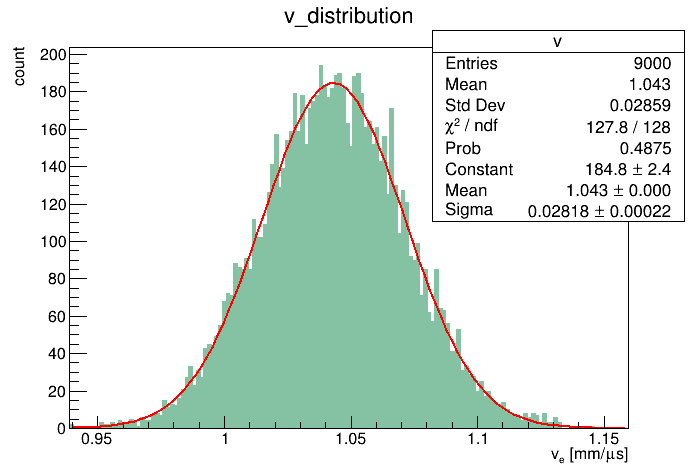}
  \includegraphics[width=.49\textwidth,trim=0.2cm 0cm 0cm 0cm]{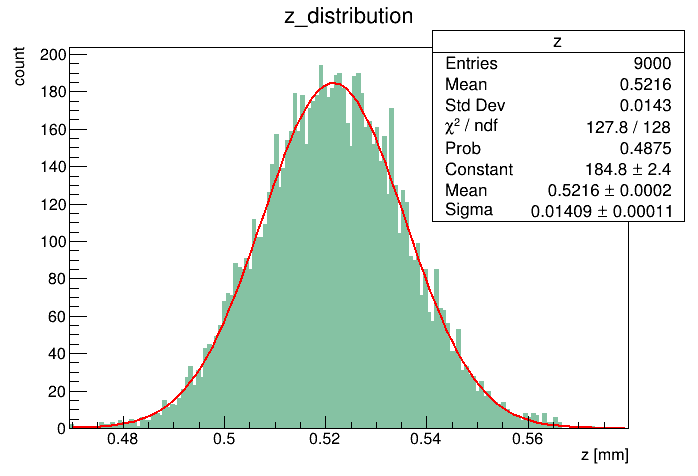}
  \includegraphics[width=.49\textwidth,trim=0.2cm 0cm 0cm 0cm]{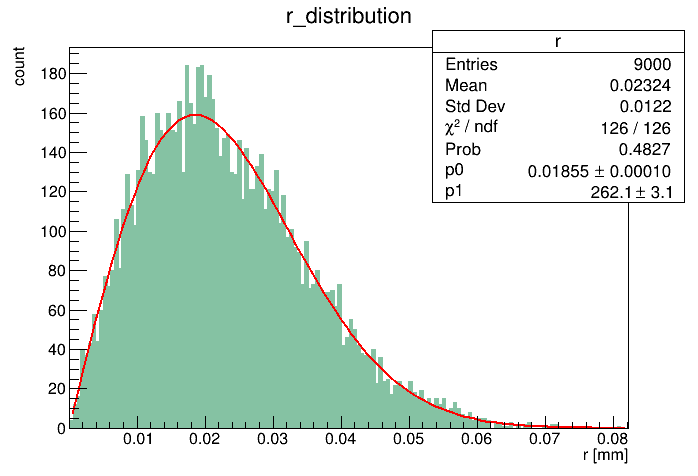}
  \caption{\label{fig:sample_distribution} Simulated electron drift velocity and position distributions under 200~V/cm electric field in liquid argon. The top plot shows the drift velocity distribution, fitted by Gaussian; the bottom-left plot shows the distribution in the longitudinal direction, fitted by Gaussian; the bottom-right plot shows the distribution in the transverse direction, fitted by equation~\eqref{eq:D_T_fit}.}
\end{figure}

Figure~\ref{fig:sample_distribution} illustrates an instance of the simulated electron drift velocity, as well as longitudinal and transverse distributions of an electron traveling in liquid argon, with an electric field strength set at 200~V/cm. 
The $\chi ^{2}/ndf$ values demonstrate a reasonable agreement between simulation and the form of the transition probability from the Langevin equation, validating the rationale for extracting parameters from distribution fitting.

Similar to what is shown in figure~\ref{fig:sample_distribution}, transportation parameters with different electric fields have been extracted from the fitting of the simulation data generated by the Monte Carlo simulation tool following the same procedure. 
The drift velocity, distribution in the longitudinal direction, and distribution in the transverse direction can be summarized by the following three equations:
\begin{subequations}
  \label{eq:fit_value}
  \begin{align}
    \label{eq:fit_value:1}
    v_{e} & = \mu_{v}, \\
    \label{eq:fit_value:2}
    D_{L} & = \frac{\sigma^{2}_{L}}{2t}, \\
    \label{eq:fit_value:3}
    D_{T} & = \frac{\sigma^{2}_{T}}{2t}.
  \end{align}
\end{subequations}

\begin{figure}[htbp]
  \centering
  \subfigure[Velocity in liquid argon]{\label{fig:results_1}
  \includegraphics[width=.49\textwidth,trim=1cm 0.5cm 0.5cm 1cm]{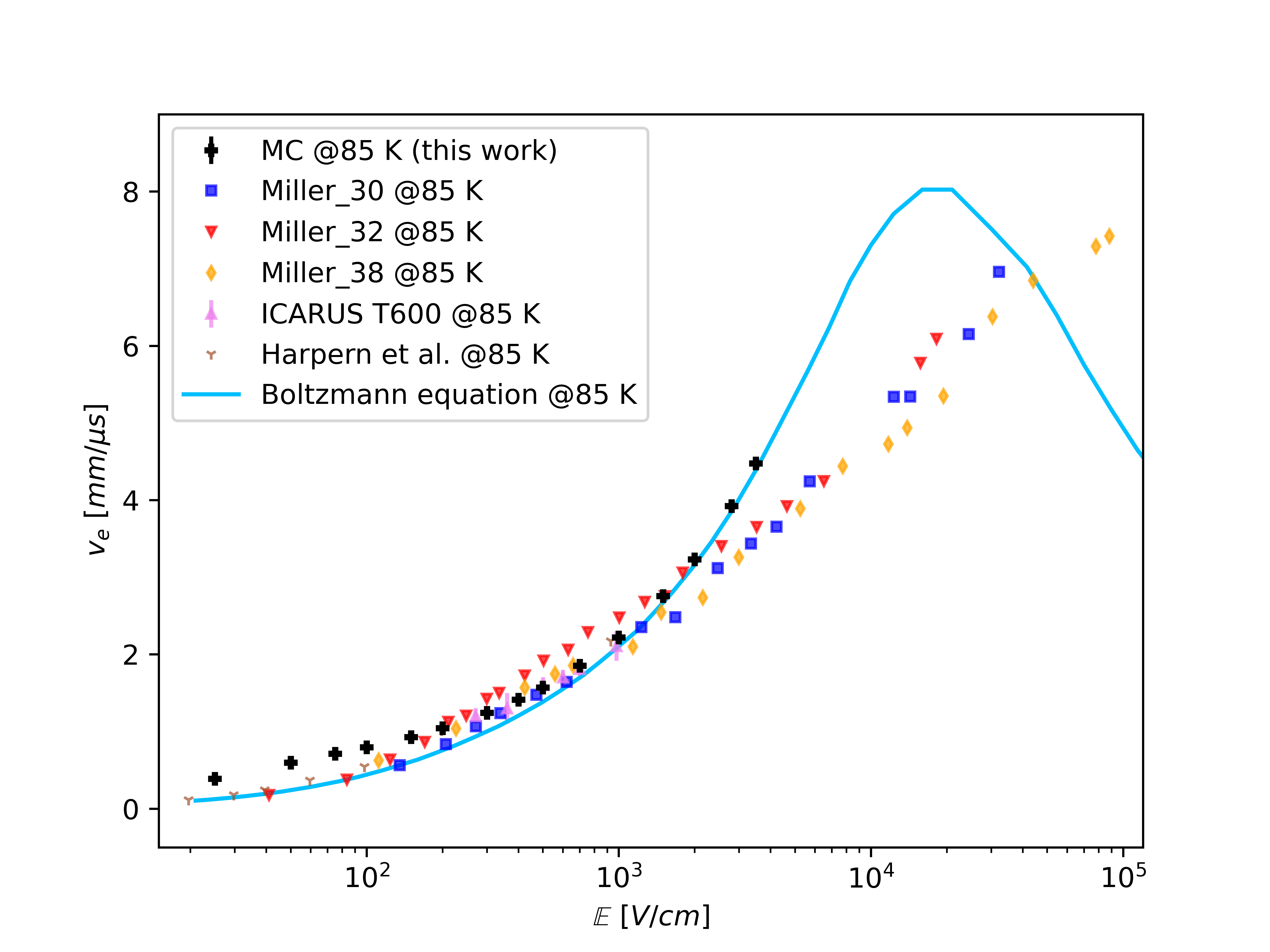}}
  \subfigure[Velocity in liquid xenon]{\label{fig:results_3}
  \includegraphics[width=.49\textwidth,trim=0.5cm 0.5cm 1cm 1cm]{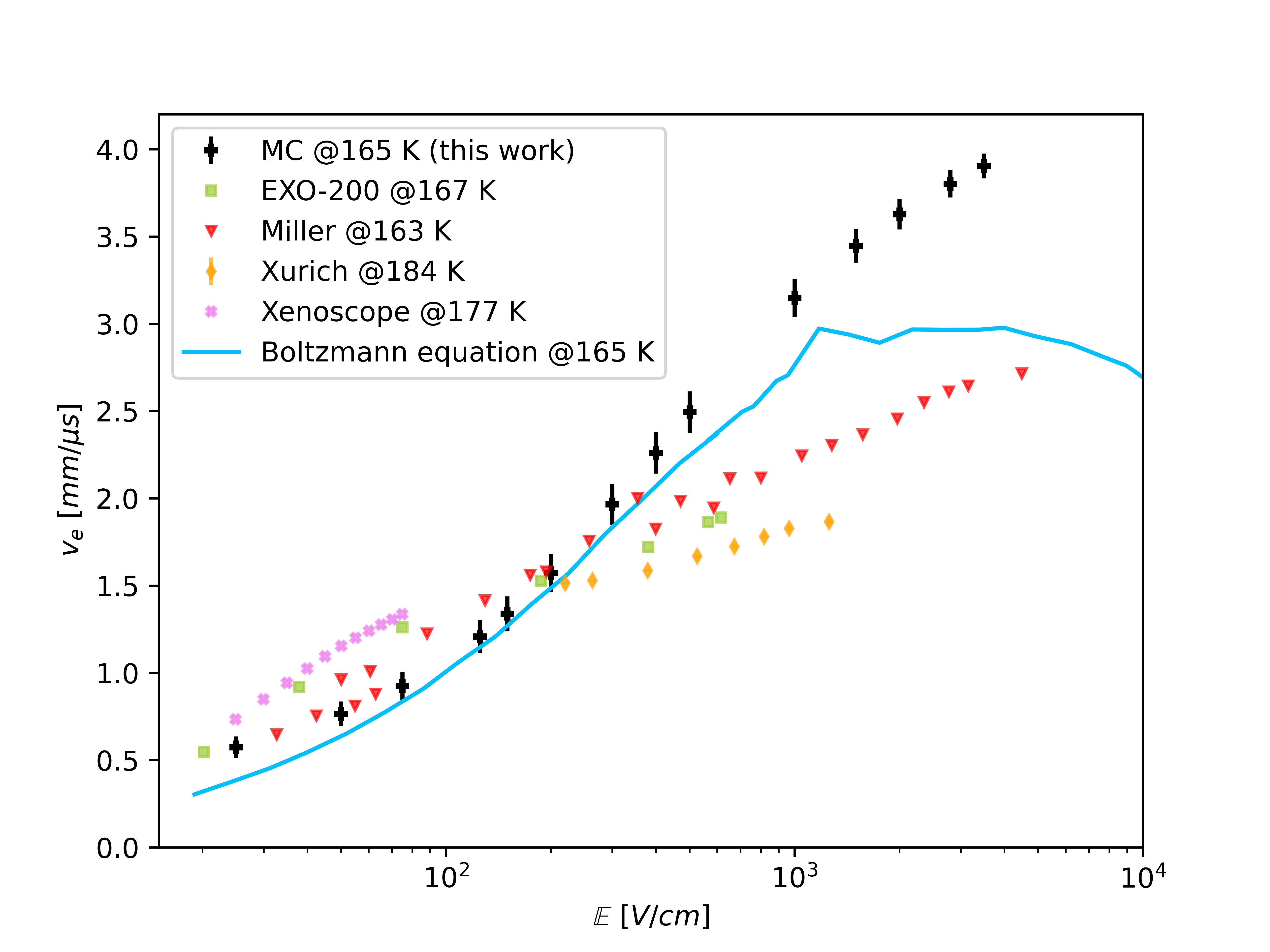}}
  \hfill
  \subfigure[Longitudinal diffusion in liquid argon]{\label{fig:results_2}
  \includegraphics[width=.49\textwidth,trim=1cm 0.5cm 0.5cm 0.7cm]{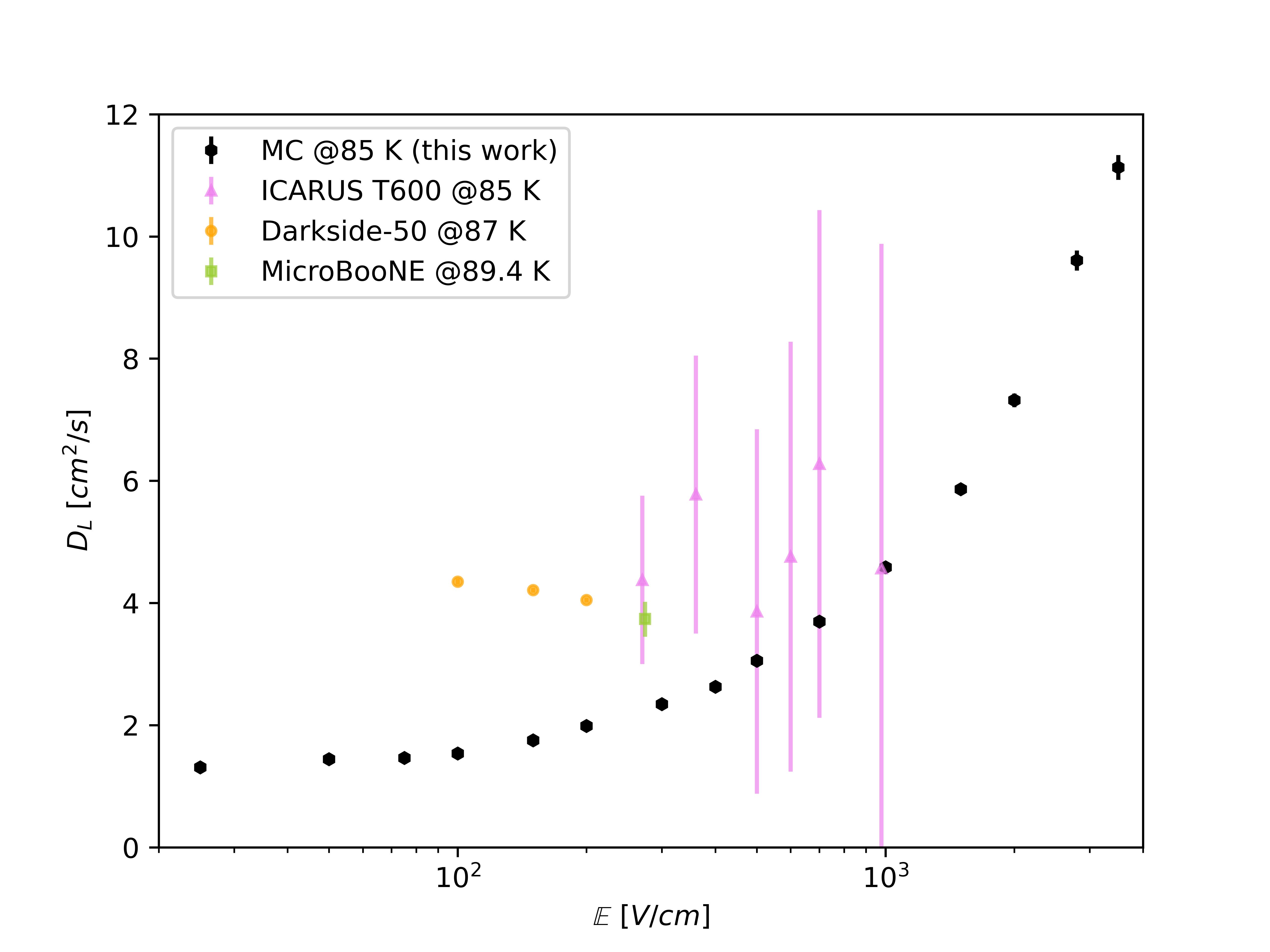}}
  \subfigure[Longitudinal diffusion in liquid xenon]{\label{fig:results_4}
  \includegraphics[width=.49\textwidth,trim=0.5cm 0.5cm 1cm 0.7cm]{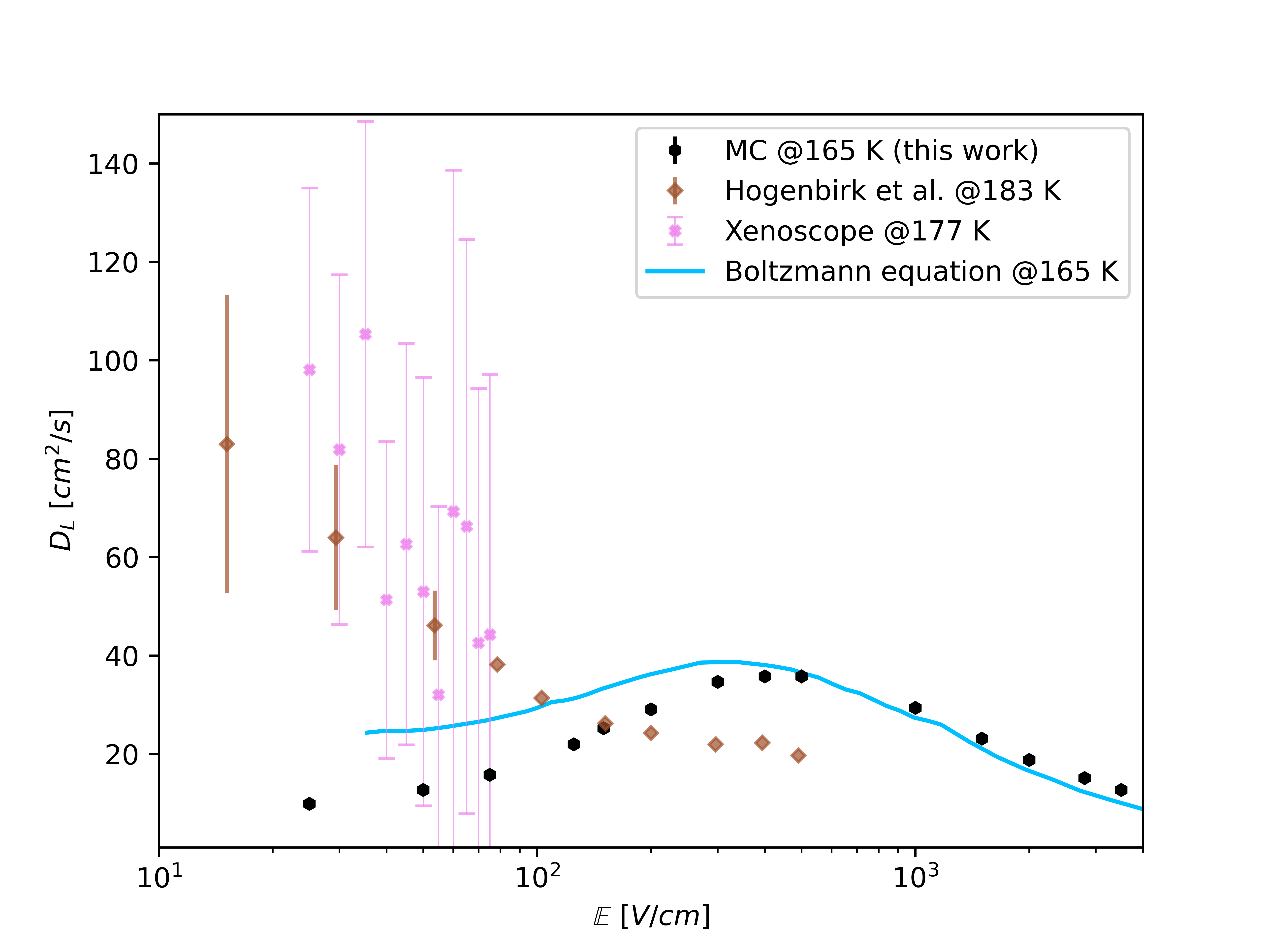}}
  \hfill
  \subfigure[Transverse diffusion in liquid argon]{\label{fig:results_5}
  \includegraphics[width=.49\textwidth,trim=1cm 0.5cm 0.5cm 0.7cm]{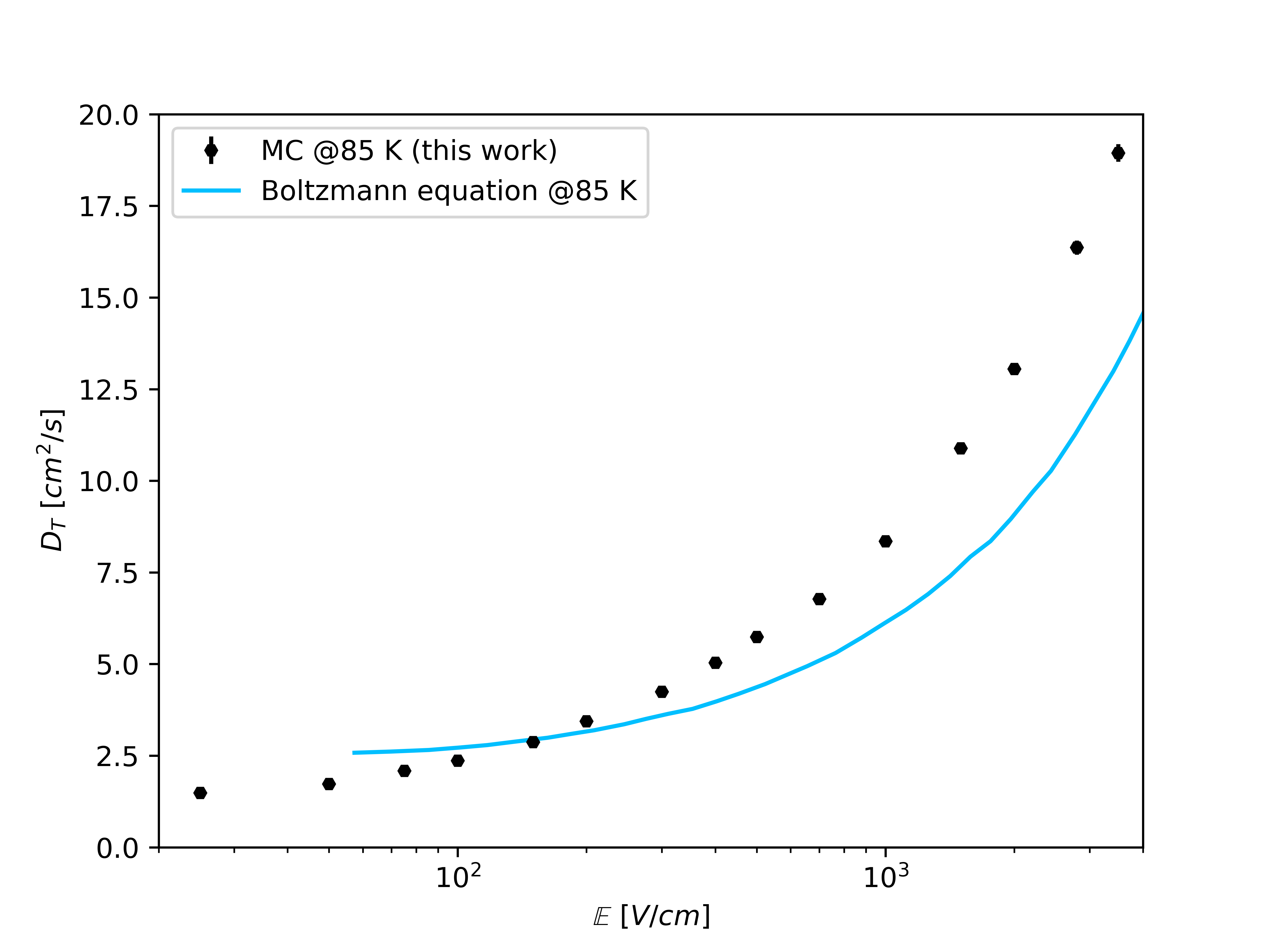}}
  \subfigure[Transverse diffusion in liquid xenon]{\label{fig:results_6}
  \includegraphics[width=.49\textwidth,trim=0.5cm 0.5cm 1cm 0.7cm]{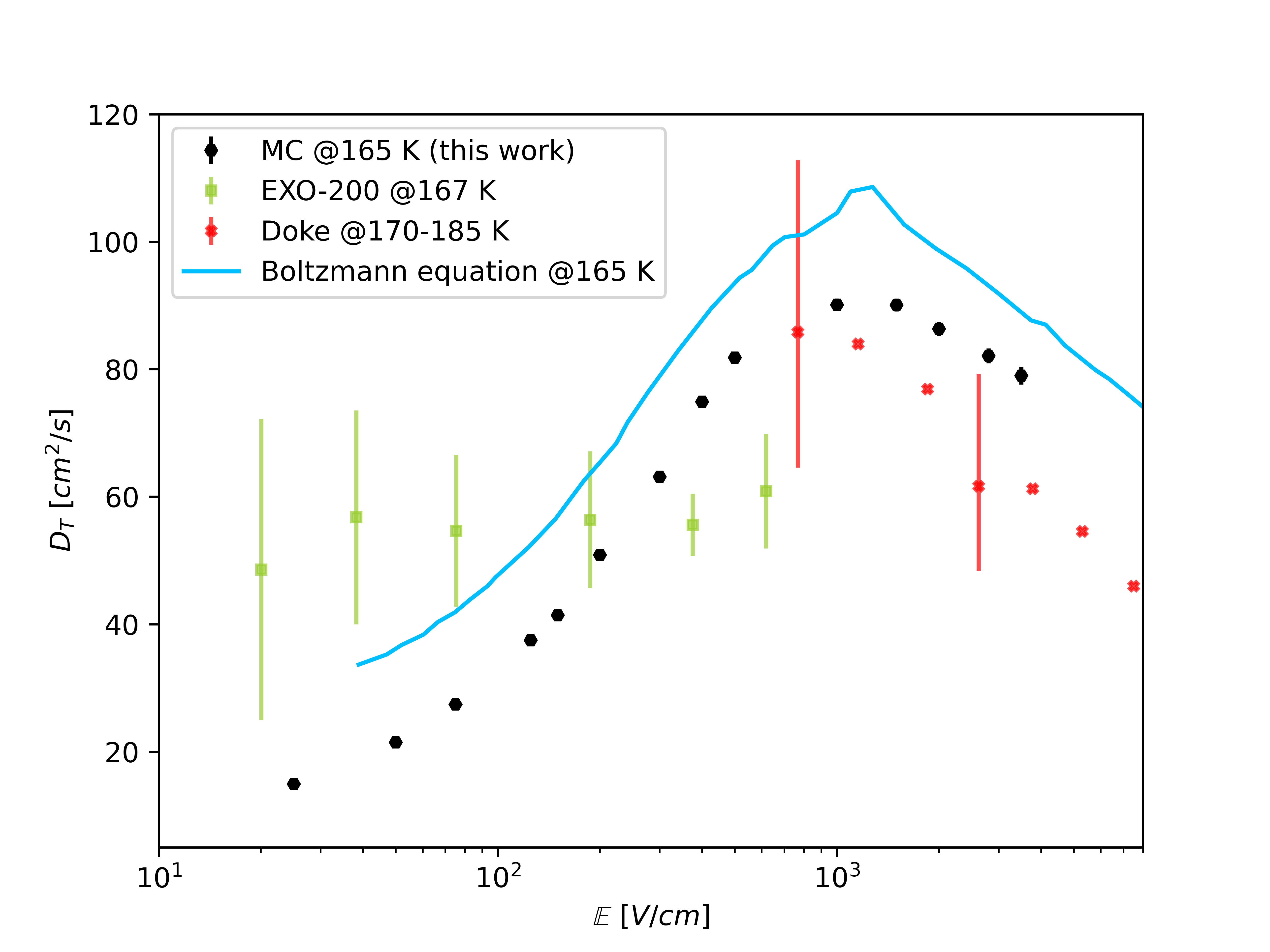}}
  \caption{\label{fig:results} Monte Carlo simulation results of the electron transportation parameters as a function of electric fields in liquid argon and liquid xenon. The top-left plot indicates the drift velocity in liquid argon at 85~K, with experimental measurements from~\cite{ICARUS, Miller, Halpern}; the middle-left plot indicates the longitudinal diffusion coefficients in liquid argon at 85~K, with experimental measurements from~\cite{ICARUS,DS,MB}; the bottom-left plot indicates the transverse diffusion coefficients in liquid argon at 85~K. The top-right plot indicates the drift velocity in liquid xenon at 165~K, with experimental measurements from~\cite{exo200e, Miller, Xurich, xenoscope}; the middle-right plot indicates the longitudinal diffusion coefficients in liquid xenon at 165~K, with experimental measurements from~\cite{Hogenbirk, xenoscope}; the bottom-right plot indicates the transverse diffusion coefficients in liquid xenon at 165~K, with experimental measurements from~\cite{exo200e,Doke,Doke2,Doke3}. Solutions of the multi-term Boltzmann equation are also plotted as a reference.}
\end{figure}

Figure~\ref{fig:results} displays the simulation results with the developed Monte Carlo simulation tool, comparing with several different experimental measurements of the electron transportation parameters in liquid argon and liquid xenon, respectively. 
A calculation from a multi-term Boltzmann equation~\cite{Boyle,Boyle2} is also included as a reference.

Drift velocity simulated from the developed Monte Carlo simulation tool, as depicted in figure~\ref{fig:results_1} and figure~\ref{fig:results_3}, aligns well with experimental measurements in both liquid argon and liquid xenon in the low to moderate electric field regions. 
In liquid argon, the simulation explains the data well in the region between 200~V/cm and 2000~V/cm, while in liquid xenon, this well-fitted region moves to somewhere below 300~V/cm. 

The remaining four plots in figure~\ref{fig:results} illustrate diffusion coefficients simulated using the developed Monte Carlo simulation tool.
For transverse diffusion in liquid argon, as there is no conceivable experimental data available, only Monte Carlo simulation results are presented, compared to calculations from the multi-term Boltzmann equation.
In the case of longitudinal diffusion in liquid argon and diffusion in liquid xenon, there are disagreements in trends between the Monte Carlo simulation and experimental measurements in the low electric field ranges, specifically from 100 V/cm to 500 V/cm in Figure~\ref{fig:results_2}, below 500 V/cm in Figure~\ref{fig:results_4}, and below 700 V/cm in Figure~\ref{fig:results_6}.
Presently, there is no satisfactory explanation for this disagreement. However, owing to the considerable size of the error bars in the experimental measurements, there isn't enough information to judge how well the simulation aligns with the data.
More measurements are definitely needed to further precisely explain the diffusion coefficients of the electron transport in noble liquids.

In general, for both drift velocity and diffusion coefficient, we posit several factors that could potentially contribute to or cause the discrepancies between the simulation and experimental measurements.
First, in the specific case of drift velocity in liquid argon at electric fields below 200~V/cm, the discrepancy may arise from the utilization of inaccurate energy transfer parameter, $\omega$.
Second, the accuracy of calculating the cross-section of coherent scattering between electrons and atoms needs further refinement, particularly under high electric fields. This may be a primary factor contributing to the discrepancies in drift velocity under high electric fields in both liquid argon and liquid xenon.
Third, the structure factor $S(\boldsymbol{\kappa})$ is sensitive to minor temperature changes within a certain range~\cite{Miller}. This sensitivity could introduce additional inaccuracies, given that experimental measurements are typically conducted at temperatures that may not be precisely determined.
Lastly, the influence of impurities within the noble liquids is not considered in this study, which could be a significant factor contributing to the discrepancies between simulations and experimental measurements.

\section{Conclusions and prospects}

We have derived a kinematic equation for ionized thermal electrons originating from Brownian movement, designed for simulating electron transportation processes in noble liquids. 
The developed Geant4-based Monte Carlo simulation tool comprehensively models the electron's movement under an electric field and its scattering process with atoms during travel. 
This includes simulation parameters such as drift velocity, longitudinal diffusion coefficient, and transverse diffusion coefficient for varying electric fields.

Comparison between the simulation and experimental measurements reveals reasonable agreements in drift velocity for electric fields ranging from 200~V/cm to 2000~V/cm in liquid argon and below 300~V/cm in liquid xenon. 
However, due to imprecise experimental measurements of the diffusion coefficient, the efficiency and precision of the developed simulation tool are yet to be conclusively determined on this aspect. 
Further precision in measurements is recommended to enhance judgment and improvement of the tool and gain a better understanding of its microscopic mechanisms.

Looking ahead, there are prospects for enhancing the simulation tool based on specific application demands.
The observed deviation in drift velocity in low-field region may be attributed to inaccuracies in energy transfer sampling. Future enhancements may involve utilizing high-order moments of energy transfer $\omega$ through methods such as long-time approximation~\cite{Sears, Sears1} or Gram-Charlier expansion~\cite{Sears2, Sears3, van_Well}, with plans to integrate these methods into the package.
Furthermore, we intend to consider interactions between electrons and impurities during drift to make the simulations more realistic in the experimental environment.
In subsequent studies, the non-uniformity of the electric field near the wire gate or cathode of the TPC is an important consideration. Our Monte Carlo package is equipped to apply an electric field map for electron transport studies, allowing us to assess the influence of field non-uniformity on signal formation.
Additionally, there are plans to integrate electron-ion recombination and attachment mechanisms into this package in future developments.

\section*{Source code availability}

The source code could be made available on request.

\acknowledgments

This work is supported by the National Key Research and Development Project of China, Grant No. 2022YFA1602001.

The authors also express gratitude to Dr. Gregory James Boyle from James Cook University, who provides us with the coherent electron-atom scattering cross-section data for completing this work.


\begin{thebibliography}{99}

\bibitem{darkside50h}
P.~Agnes et al., \emph{DarkSide-50 532-day dark matter search with low-radioactivity argon}, \href{https://doi.org/10.1103/PhysRevD.98.102006}{\emph{Phys. Rev. D} {\bf 98} (2018) 102006}.

\bibitem{darkside50l}
P.~Agnes et al., \emph{Search for low-mass dark matter WIMPs with 12 ton-day exposure of DarkSide-50}, \href{https://doi.org/10.1103/PhysRevD.107.063001}{\emph{Phys. Rev. D} {\bf 107} (2023) 063001}.

\bibitem{xenonnt}
E.~Aprile et al., \emph{First Dark Matter Search with Nuclear Recoils from the XENONnT Experiment}, \href{https://doi.org/10.1103/PhysRevLett.131.041003}{\emph{Phys. Rev. Lett.} {\bf 131} (2023) 041003}.

\bibitem{LZ}
J.~Aalbers et al., \emph{First Dark Matter Search Results from the LUX-ZEPLIN (LZ) Experiment}, \href{https://doi.org/10.1103/PhysRevLett.131.041002}{\emph{Phys. Rev. Lett.} {\bf 131} (2023) 041002}.

\bibitem{pandax4t}
Y.~Meng et al., \emph{Dark Matter Search Results from the PandaX-4T Commissioning Run}, \href{https://doi.org/10.1103/PhysRevLett.127.261802}{\emph{Phys. Rev. Lett.} {\bf 127} (2021) 261802}.

\bibitem{protodune}
B.~Abi et al., \emph{First results on ProtoDUNE-SP liquid argon time projection chamber performance from a beam test at the CERN Neutrino Platform}, \href{https://doi.org/10.1088/1748-0221/15/12/P12004}{\emph{JINST} {\bf 15} (2020) P12004}.

\bibitem{exo200r}
G.~Anton et al., \emph{Search for Neutrinoless Double-
$beta$ Decay with the Complete EXO-200 Dataset}, \href{https://doi.org/10.1103/PhysRevLett.123.161802}{\emph{Phys. Rev. Lett.} {\bf 123} (2019) 161802}.

\bibitem{exo200e}
J.~B.~Albert et al., \emph{Measurement of the drift velocity and transverse diffusion of electrons in liquid xenon with the EXO-200 detector}, \href{https://doi.org/10.1103/PhysRevC.95.025502}{\emph{Phys. Rev. C} {\bf 95} (2017) 025502}.

\bibitem{ICARUS}
M.~Torti et al., \emph{Electron diffusion measurements in the ICARUS T600 detector}, \href{https://doi.org/10.1088/1742-6596/888/1/012060}{\emph{J. Phys.: Conf. Ser.} {\bf 888} (2017) 012060}.

\bibitem{NEST}
M.~Szydagis et al., \emph{NEST: a comprehensive model for scintillation yield in liquid xenon}, \href{https://doi.org/10.1088/1748-0221/6/10/P10002}{\emph{JINST} {\bf 6} (2011) P10002}.

\bibitem{translate}
Z.~Beever et al., \emph{TRANSLATE - a Monte Carlo simulation of electron transport in liquid argon}, \href{https://doi.org/10.1016/j.cpc.2023.109056}{\emph{Comput. Phys. Commun.} {\bf 297} (2024) 109056}.

\bibitem{Van_Hove}
L.~Van~Hove, \emph{Correlations in Space and Time and Born Approximation Scattering in Systems of Interacting Particles}, \href{https://doi.org/10.1103/PhysRev.95.249}{\emph{Phys. Rev.} {\bf 95} (1954) 249}.

\bibitem{geant4}
S.~Agostinelli et al., \emph{Geant4—a simulation toolkit}, \href{https://doi.org/10.1016/S0168-9002(03)01368-8}{\emph{Nucl. Instrum. Methods Phys. Res. Sect. A} {\bf 506} (2003) 250}.

\bibitem{ST}
Manuel~Osvaldo~Cáceres, \emph{Non-equilibrium statistical physics with application to disordered systems}, Springer Cham (2017).

\bibitem{Levy}
Andreas~E.~Kyprianou, \emph{The L{\'e}vy--It{\^o} decomposition and path structure}, Springer, Berlin, Heidelberg (2014).

\bibitem{Langevin}
Don~S.~Lemons, and Anthony~Gythiel, \emph{Paul Langevin's 1908 paper “On the Theory of Brownian Motion” ["Sur la théorie du mouvement brownien," C. R. Acad. Sci. (Paris) 146, 530-533 (1908)]}, \href{https://doi.org/10.1119/1.18725}{\emph{Am. J. Phys.} {\bf 65} (1997) 1079-1081}.

\bibitem{Ito}
Philip~E.~Protter, \emph{Stochastic Integration and Differential Equations}, \href{https://doi.org/10.1007/978-3-662-10061-5}{Springer Berlin, Heidelberg (2005)}.

\bibitem{Lekner}
J.~Lekner, \emph{Motion of Electrons in Liquid Argon}, \href{https://doi.org/10.1103/PhysRev.158.130}{\emph{Phys. Rev.} {\bf 158} (1967) 130}.

\bibitem{Percus}
M.~S.~Wertheim, \emph{Exact Solution of the Percus-Yevick Integral Equation for Hard Spheres}, \href{https://doi.org/10.1103/PhysRevLett.10.321}{\emph{Phys. Rev. Lett.} {\bf 10} (1963) 321}.

\bibitem{Yarnell}
J.~L.~Yarnell et al., \emph{Structure Factor and Radial Distribution Function for Liquid Argon at 85 \ifmmode^\circ\else\textdegree\fi{}K}, \href{https://doi.org/10.1103/PhysRevA.7.2130}{\emph{Phys. Rev. A} {\bf 7} (1973) 2130}.

\bibitem{Atrazhev_Liquid_Xenon}
V.~M.~Atrazhev et al., \emph{Electron transport coefficients in liquid xenon}, \href{https://doi.org/10.1109/ICDL.2005.1490092}{\emph{Proc. IEEE Int. Conf. Dielectr. Liq.} Coimbra, Portugal (2005)}.

\bibitem{Boyle2}
G.~J.~Boyle et al., \emph{Ab initio electron scattering cross-sections and transport in liquid xenon}, \href{https://doi.org/10.1088/0022-3727/49/35/355201}{\emph{J. Phys. D: Appl. Phys.} {\bf 49} (2016) 355201}.

\bibitem{Boyle}
G.~J.~Boyle et al., \emph{Electron scattering and transport in liquid argon}, \href{https://doi.org/10.1063/1.4917258}{\emph{J. Chem. Phys.} {\bf 142} (2015) 154507}.

\bibitem{G4_1}
J.~Allison et al., \emph{Recent developments in Geant4}, \href{https://doi.org/10.1016/j.nima.2016.06.125}{\emph{Nucl. Instrum. Methods Phys. Res. Sect. A} {\bf 835} (2016) 186}.

\bibitem{G4_2}
J.~Allison et al., \emph{Geant4 developments and applications}, \href{https://doi.org/10.1109/TNS.2006.869826}{\emph{IEEE Trans. Nucl. Sci.} {\bf 53} (2006) 270}.

\bibitem{Rahman}
A.~Rahman, K.~S.~Singwi and A.~Sj{\"o}lander, \emph{Theory of Slow Neutron Scattering by Liquids. I}, \href{https://doi.org/10.1103/PhysRev.126.986}{\emph{Phys. Rev.} {\bf 126} (1962) 986}.

\bibitem{Miller}
L.~S.~Miller, S.~Howe and W.~E.~Spear, \emph{Charge Transport in Solid and Liquid Ar, Kr, and Xe}, \href{https://doi.org/10.1103/PhysRev.166.871}{\emph{Phys. Rev.} {\bf 166} (1968) 871}.

\bibitem{Halpern}
B.~Halpern et al., \emph{Drift Velocity and Energy of Electrons in Liquid Argon}, \href{https://doi.org/10.1103/PhysRev.156.351}{\emph{Phys. Rev.} {\bf 156} (1967) 351}.

\bibitem{DS}
P.~Agnes et al., \emph{Electroluminescence pulse shape and electron diffusion in liquid argon measured in a dual-phase TPC}, \href{https://doi.org/10.1016/j.nima.2018.06.077}{\emph{Nucl. Instrum. Methods Phys. Res. Sect. A} {\bf 904} (2018) 23}.

\bibitem{MB}
P.~Abratenko et al., \emph{Measurement of the longitudinal diffusion of ionization electrons in the MicroBooNE detector}, \href{https://doi.org/10.1088/1748-0221/16/09/P09025}{\emph{JINST} {\bf 16} (2021) P09025}.

\bibitem{Xurich}
L.~Baudis et al., \emph{A dual-phase xenon TPC for scintillation and ionisation yield measurements in liquid xenon}, \href{https://doi.org/10.1140/epjc/s10052-018-5801-5}{\emph{Eur. Phys. J. C} {\bf 78} (2018) 351}.

\bibitem{xenoscope}
L.~Baudis et al., \emph{Electron transport measurements in liquid xenon with Xenoscope, a large-scale DARWIN demonstrator}, \href{https://doi.org/10.1140/epjc/s10052-023-11823-1}{\emph{Eur. Phys. J. C} {\bf 83} (2023) 717}.

\bibitem{Hogenbirk}
E.~Hogenbirk et al., \emph{Field dependence of electronic recoil signals in a dual-phase liquid xenon time projection chamber}, \href{https://doi.org/10.1088/1748-0221/13/10/P10031}{\emph{JINST} {\bf 13} (2018) P10031}.

\bibitem{Doke2}
T.~Doke, \emph{Recent developments of liquid xenon detectors}, \href{https://doi.org/10.1016/0029-554X(82)90621-8}{\emph{Nucl. Instrum. Methods Phys. Res.} {\bf 196} (1982) 87}.

\bibitem{Doke3}
T.~Doke, \emph{Fundamental properties of liquid argon, krypton and xenon as radiation detectro media}, {\emph{Portgal. Phys.} {\bf 12} (1981) 9}.

\bibitem{Doke}
E.~Aprile and T.~Doke, \emph{Liquid xenon detectors for particle physics and astrophysics}, \href{https://doi.org/10.1103/RevModPhys.82.2053}{\emph{Rev. Mod. Phys.} {\bf 82} (2010) 2053}.

\bibitem{Sears}
V.~F.~Sears, \emph{Continued fraction representation for slow neutron scattering}, \href{https://doi.org/10.1139/p69-023}{\emph{Can. J. Phys.} {\bf 47} (1969) 199}.

\bibitem{Sears1}
V.~F.~Sears, \emph{Continued fraction representation for slow neutron scattering. II}, \href{https://doi.org/10.1139/p70-079}{\emph{Can. J. Phys.} {\bf 48} (1970) 616}.

\bibitem{Sears2}
V.~F.~Sears, \emph{High-Energy Neutron Scattering from Liquid ${\mathrm{He}}^{4}$}, \href{https://doi.org/10.1103/PhysRev.185.200}{\emph{Phys. Rev.} {\bf 185} (1969) 200}.

\bibitem{Sears3}
V.~F.~Sears, \emph{High-Energy Neutron Scattering from Liquid ${\mathrm{He}}^{4}$. II. Interference and Temperature Effects}, \href{https://doi.org/10.1103/PhysRevA.1.1699}{\emph{Phys. Rev. A} {\bf 1} (1970) 1699}.

\bibitem{van_Well}
A.~A.~van~Well and L.~A.~de~Graaf, \emph{Density fluctuations in liquid argon. II. Coherent dynamic structure factor at large wave numbers}, \href{https://doi.org/10.1103/PhysRevA.32.2384}{\emph{Phys. Rev. A} {\bf 32} (1985) 2384}.

\end{thebibliography}
\end{document}